\documentclass[preprint,aps,showpacs]{revtex4}

\usepackage{amsmath,amssymb}
\usepackage{graphicx}
\usepackage{placeins}
\usepackage{epsfig}

\newcommand{\be}{\begin{equation}}
\newcommand{\ee}{\end{equation}}
\newcommand{\ba}{\begin{eqnarray}}
\newcommand{\ea}{\end{eqnarray}}
\newcommand{\1}{\frac}

\newcommand{\pa}{\partial}

\newcommand{\la}{\label}                                                                                                

\renewcommand{\inf}{\infty}

\begin{document}

\title{\Large \bf Cosmology with decaying tachyon matter}

\author{A. Das${}^{*}$, Shashikant Gupta${}^{\dagger}$, Tarun Deep Saini${}^{\dagger}$ and Sayan Kar${}^*$}

\affiliation{\rm Department of Physics and Meteorology and 
CTS \\ Indian Institute of Technology Kharagpur 721302, India}

\email{anupam@phy.iitkgp.ernet.in, sayan@cts.iitkgp.ernet.in}

\affiliation{\rm Department of Physics, Indian Institute of 
Science, Bangalore 560012 India}

\email{shashikant@physics.iisc.ernet.in, tarun@physics.iisc.ernet.in}

\begin{abstract}
We  investigate the case of a homogeneous tachyon field coupled to gravity 
in a spatially flat Friedman-Robertson-Walker spacetime. Assuming the field 
evolution to be exponentially decaying with time we solve the field equations 
and show that, under certain conditions, the scale factor represents an 
accelerating universe, following a phase of decelerated expansion. 
We make use of a model of dark energy (with $p=-\rho$) and dark matter
($p=0$) where a single scalar field (tachyon) governs the dynamics of
both the dark components. We show that this model fits the current 
supernova data as well as the canonical $\Lambda$CDM model. We give
the bounds on the parameters allowed by the current data.

\end{abstract}


\maketitle

\newpage

\section{Introduction}

The observational fact that the present universe is accelerating  {\cite{accl}}
has led to several theoretical speculations  
about the nature of the acceleration and the cause behind it. No completely 
consistent model exists till date,
though a plethora of them with intrinsic advantages and discrepancies
have been proposed (for a sample of these in recent literature see {\cite{acclth}}). 
The cause of the acceleration is attributed to
the presence of a mysterious entity named {\em dark energy} which
dominates (is about 70 percent) the matter--energy content of the
present day universe {\cite{dark}}. 
In order to drive the acceleration the Dark energy
needs negative pressure. For an equation of state of the form $p=w \rho$
observations place bounds on $w$ : $-1.33 \leq w \leq -0.8$. Thus models
respecting $p=w \rho$ must be consistent with these bounds. Among many, the
simplest is Einstein's old idea of an universe with a cosmological constant, 
for which $w=-1$. In a more general scenario the case of a time dependent $w(t)$ has also been looked at to some extent.  

The majority of dark energy models involve the temporal dynamics of
one or more scalar fields with varied actions and with or without
a scalar potential {\cite{scalarfield}. A sub-class of these scalar
field models includes exotic models where the kinetic energy term
may become negative (the phantom) {\cite{phantom}}
 or the scalar action may be of
a non--standard form (the tachyon condensate). The latter arises 
in the context of theories of unification such as superstring theory
{\cite{tachyon}}. 
In this article, we shall concentrate on the the cosmological
consequences of the class of models where the scalar tachyon 
condensate appears in the matter part of the Einstein equations.

Let us first briefly review the origins of the effective action for the 
tachyon condensate. It is known from the early days of string theory
that the spectrum of strings do contain a negative mass squared 
vibration called the tachyon. For strings attached to Dirichlet (D)
branes such tachyonic modes reflect D-brane instability. The
low energy effective action for the scalar tachyon field 
around the tachyon vacuum thus 
contains a tachyon potential which has an unstable maximum resulting
in a negative mass square value in the mass term. The scalar tachyon 
rolls down from the potential maximum to a more stable configuration
thereby acquiring a stable vacuum expectation value. This is the
tachyon condensate. The questions in cosmology are : (i) Can such a rolling
tachyon drive the acceleration of the universe? (ii) Can the tachyon
condensate be of use in describing all that is described in cosmology
using scalar fields (inflation, dark matter etc)? A distinct advantage
of using the scalar tachyon to understand cosmology is the fact that
we do not need to answer the embarrassing question `where does the 
scalar field come from?' Note that this is reminiscent of the oft-quoted 
advantage of string (dilaton) cosmology too {\cite{dilaton}}. 
      
There have been several articles in the literature where the
cosmological consequences of the tachyon condensate has been
dealt with {\cite{gibbons1}}. It was shown in {\cite{sen1}} that the effective tachyon action 
can lead to an energy--momentum equivalent to a gas with zero pressure 
(dust) but nonzero energy density (modulo some assumptions
and re--definitions). 
It has also been noticed in {\cite{padmanabhan1} that, modulo some
assumptions, the tachyon field can also be made use of in describing
dark matter at galactic scales. In addition, the energy momentum tensor
of the tachyon condensate can be split into two parts--one with zero
pressure (dark matter) and another with $p=-\rho$. This facilitates
the description of dark energy and dark matter using a single scalar
field--a fact we shall test in detail for our model in this article.
Tachyons and their dynamics in the context of string theory 
have been reviewed in {\cite{tachyon}}.

Most of the work on tachyon cosmology assumes a tachyon field varying linearly with time. In addition, the choice of the tachyon potential
is also assumed to be either exponential (decaying) or inverse
quadratic \cite{padmanabhan2}. Alternative potentials satisfying the
general features (derived from string field theory) have also been
proposed and analyzed {\cite{otherpotentials}}. In our work, we shall
assume the tachyon field to be decaying exponentially with time. We will see 
that the scale factor can have an accelerating behavior for certain values
of the constants which appear in our solution. The tachyon potential for such  
scenario does satisfy the general requirements. In addition, we find different
types of solutions with the exponentially decaying tachyon field. Finally, we use Supernova data to constrain the parameters that appear in our solution.

We choose to work with units $c=1$ though, wherever
necessary we go back to the actual units and dimensions.

\section{The Analysis}
The action for the homogeneous tachyon condensate coupled to gravity is 
known to be of the form {\cite{sen1}}
\be
S=S_{\rm Grav}+S_{\rm Tachyon}=\int \sqrt{-g}\,d^4x \, \left [\1{R}{2k} -
  V(T)(1+g^{\mu\nu}\pa_\mu T \pa_\nu T)^{\frac{1}{2}} \right]\,\,, {\label{action}}
\ee
where $k=8\pi G$ and $V(T)$ is the tachyon potential. We choose the tachyon field to be 
time dependent $T=T(t)$ in a spatially flat 3+1 dimensional FRW background. The line element is 
\be
ds^2=-dt^2+a^2(t)\left (dx^2+dy^2+dz^2 \right )\,\,,
\ee 
where $a(t)$ is the scale factor. The exact form of the potential is not
known. However as we argued above, (i) it should have an unstable maximum at
the origin and (ii) should decay exponentially to zero as the field goes to
infinity. Moreover, the slope of the potential should be negative for $T>0$
{\cite{steer}}. Assuming the field stress energy tensor of the form, ${\rm
  diag}\left (\rho, p,p,p \right )$, the energy density and pressure obtained from eqn (\ref{action}) are as follows:
\begin{eqnarray}
\rho(T)=\1{V(T)}{\sqrt{1-\dot T^2}}\,\,, \\
p(T)=-V(T){\sqrt{1-\dot T^2}}\,\,.
\end{eqnarray}

Ignoring the cosmological constant the two independent components of the Einstein field equation give:
\ba
3\left(\frac{\dot a}{a} \right )^2 = \frac{kV(T)}{\sqrt{1-\dot T^2}} {\label {1st}}\\
\left ({\frac{\dot a}{a}} \right )^2+ 2\,\frac{\ddot a}{a} = kV(T)\sqrt{1-\dot T^2} {\label{2nd}}\,\,,
\ea
For the three unknown functions ($V(T), a(t)$ and $T(t)$) 
we have two equations and to solve for these we need to choose one of them 
suitably. Since no initial or boundary conditions are imposed, some arbitrary 
parameters are expected to appear in the solutions but we will see that these 
will not be completely arbitrary in order to give consistent, physically 
acceptable solutions. Also these parameters will be finally estimated by 
making a fit with observed supernova data. Depending on one's choice of field 
or potential or the scale factor, a large class of consistent solutions can 
be constructed. (e.g. {\cite{fein}} discusses exact solutions starting with 
chosen form of inverse square potential, and {\cite{steer}} discusses solutions based on choice of $\dot T(t)$, whereas {\cite{sami1}} discusses solutions 
obtained from a different choice of the potential (exponentially decaying)). 
One problem that might arise in the approach we adopt, is that the 
resulting potential may not be physical. This will be revealed as we proceed 
with the calculations.  Assuming the tachyon field to be decaying exponentially with time we have
\be
T(t)= T_0 e^{-\alpha t}\,\,,
\ee
where $\alpha$ is a positive constant and $T_0$ is the value of the field 
at $t=0$. Using this form of the field in (\ref{1st}),(\ref{2nd}) and 
eliminating $V(T)$ one gets the Hubble parameter for the universe:

\be
\frac{\dot a}{a}= \frac{1}{C-\frac{3\alpha T_0^2}{4}e^{-2\alpha t}} \,\,, \la{aeqn}
\ee 
where $C$ is an arbitrary integration constant. Incidentally, 
a physically acceptable theory is obtained only when $C$ is positive. 
($C\leq0$ gives collapsing universes.) Integrating the last relation and 
using the fact that $a(t=t_0,\mathrm{present\,\,epoch})=1$ we get

\be
a(t)=\left(\frac{e^{2\alpha t} - \frac{3 \alpha T_0^2}{4C}}{e^{2\alpha t_0} - \frac{3 \alpha T_0^2}{4C}} \right)^{1/2C\alpha}\,\,,
\ee
where $t_0$ is the age of the universe in the model under discussion. To make 
calculations simpler we now define three dimensionless parameters:
\be
p=\frac{3 (\alpha T_0)^2}{4 \alpha C} \hspace{.1in} ; \hspace{.1in}
s= \frac{1}{2\alpha C} \hspace{.1in} ; \hspace{.1in}
r= e^{2\alpha t_0}\,\,.
\ee
Using these one can write
\be
a(t)=\left(\frac{e^{2\alpha t} - p}{r - p} \right)^s\,\,.
\ee
The other unknown function, the tachyon potential can be obtained from
\be
V(T)=\frac{3(1-\alpha^2 T^2)^{\frac{1}{2}}}{k(C-\frac{3\alpha}{4}T^2)^2} \,\,.
\ee
The potential conforms to the nature of other tachyon potentials 
provided the maximum of potential occurs at a higher value of the field. Note 
that in our case the tachyon field decays with time. This requirement leads 
to the condition $C\alpha<3$. In terms of the dimensionless parameters the time dependence of potential can be expressed as 

\be
V(t)=\frac{3(1-\frac{2p}{3s}e^{-2\alpha t})^{1/2}}{k(1-pe^{-2\alpha t})^2}\,\,.
\ee

The potential will be a  real function for all values of time if $\frac{2p}{3s}<1$ i.e. $p<\frac{3s}{2}$.  
However, from the expression for the scale factor we note that $a(t=0)=0$
(big bang) would imply a choice $p=1$ (and hence $s>2/3$).
These two conditions give  
the required potential for our purpose (see fig 
\ref{pot_fig},\ref{pot_time}) which like the $1/T^2$ potential is singular
at the origin (in time). 
It can be easily seen that the isotropic pressure and energy density also 
remain well behaved under these conditions. Another important feature of the theory is the deceleration parameter which is given by
\be
q=-\frac{(\ddot a/a)}{(\dot a/a)^2}=\frac{p}{s}e^{-2 \alpha t} -1\,\,,
\ee
and represents an accelerating universe at late times 
\begin{figure}[tbh!]
\includegraphics[width=7cm,height=7cm,angle=0]{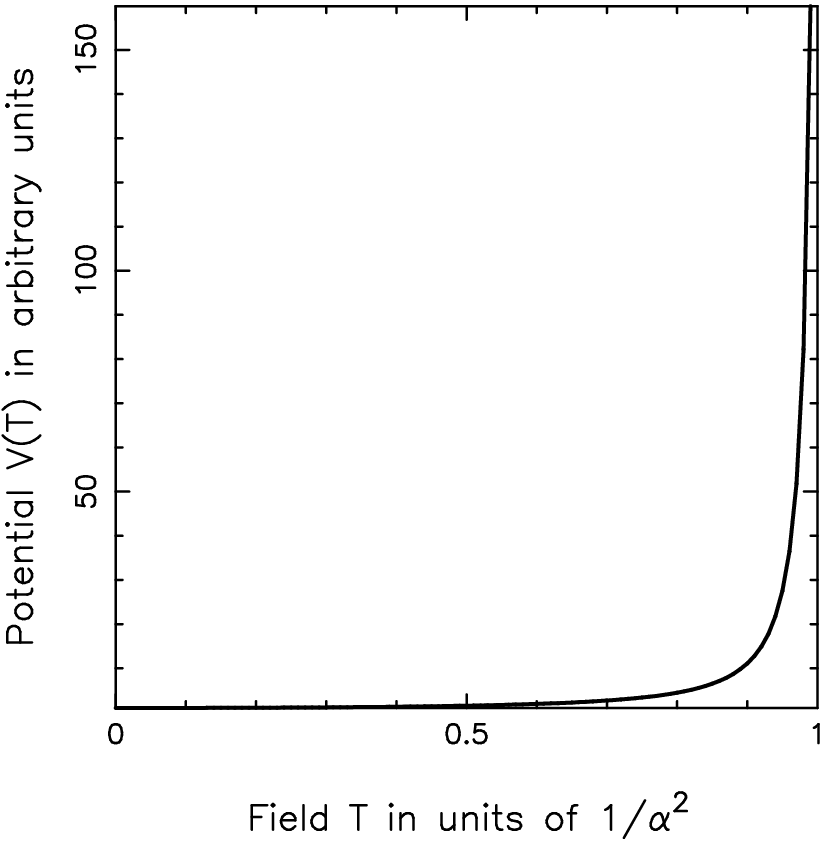}
\caption{
  We plot the potential $V(T)$ as a function of the field $T$ in arbitrary
  units. The values of the model parameters for this and the next figure are
  taken from best fit to the current supernova data given later.}
\label{pot_fig}
\end{figure}
\hfill
\begin{figure}
\includegraphics[width=7cm,height=7cm,angle=0]{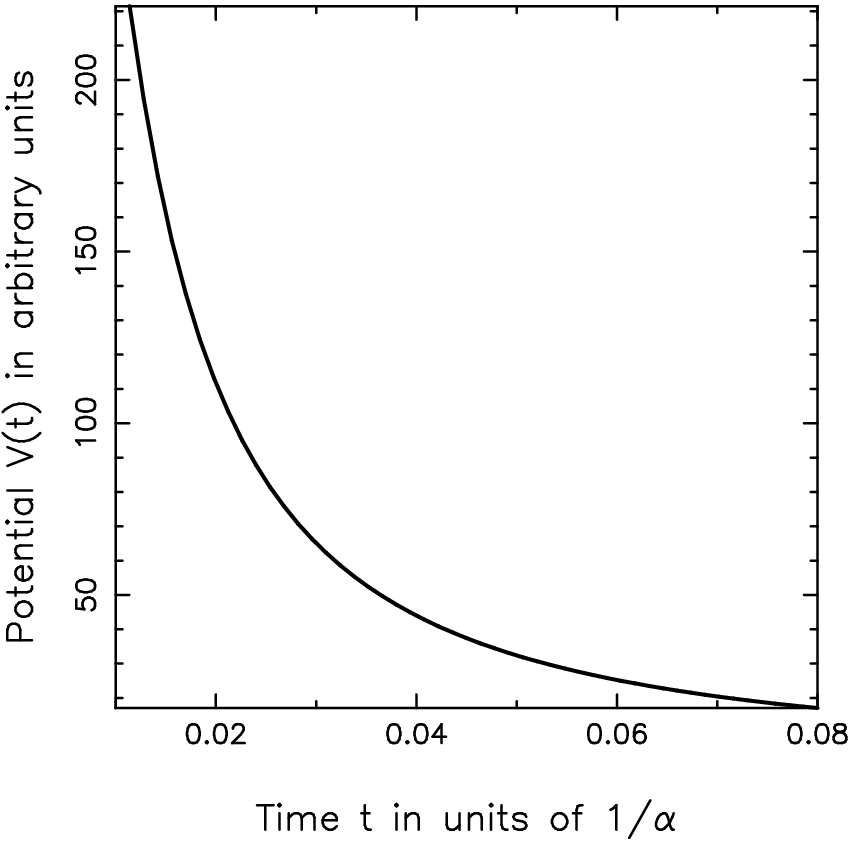} 
\caption{Potential as a function of time. The field rolls down from top of 
the potential and after infinite time interval reaches a constant value.} 
\la{pot_time}
\end{figure}

\section{Fitting the Supernova Data : Parameter Estimation}

The tachyon condensate has been proposed as a candidate for dark energy.
The above exact solution, however, does not assume anything other than
the tachyon condensate as a source in the Einstein equations. Therefore, if
the above scale factor is assumed to be correct then we have to believe in a
universe where dark energy is the sole matter source (which is unrealistic).
To remedy this deficiency we can propose the following.

(a) Add other matter fields (dark matter + $\Lambda$) and obtain the scale
factor analytically/numerically. Then obtain the unknown parameters 
by fitting  with supernova data. Finally, use these parameters to ascertain
the proportions of dark matter and dark energy.

(b) Use the split of the energy density and pressures of the
tachyon condensate into dark matter (pressure-less) and dark energy  (with $w=-1$) parts.
This goes as follows :
\begin{equation}
\rho = \rho_{DM} + \rho_{DE} \hspace{.2in} ; \hspace{.2in} p = p_{DM} + p_{DE}\,\,,
\end{equation}
where
\begin{eqnarray}
\rho_{DM} = \frac{V(T) {\dot T}^2}{\sqrt{1-{\dot T}^2}} \hspace{.2in} ;
\hspace{.2in} p_{DM} = 0 \\
\rho_{DE} = {V(T)}{\sqrt{1-{\dot T}^2}} \hspace{.2in} ; \hspace{.2in} 
p_{DE} = -\rho_{DE} \,\,.
\end{eqnarray}

In this way one should be able include the contributions of both dark matter
and dark energy.

We shall first explore the second possibility. Using the relation between
redshift and scale factor, $1+z=a_0/a(t)=1/a(t)$ one gets time-redshift
relation in the present case as:
\begin{equation}
t(z)=- \frac{1}{2\alpha}\ln\left[\frac{1}{r}\frac{(1+z)^{\frac{1}{s}}}{\left(1-\frac{p}{r}+\frac{p}{r}(1+z)^{\frac{1}{s}}\right)} \la{tr_reln}\right ]\,\,.
\end{equation}
We notice one important restriction following from the last relation. If we
impose the condition that as $t\rightarrow 0, z\rightarrow \inf$ then we must
make $p=1$. We will use this value when calculating some parameters of the
model of the universe later.  From this, one can express $\Omega_{DM}$ and
$\Omega_{DE}$ as function of redshift as follows:

\begin{eqnarray}
\Omega_{DM}(z)&=& \dot{T}^2 = T_0^2\alpha^2\,\exp \left(-2\alpha t(z)\right) \la{odm}\\
\Omega_{DE}(z) &=& 1 - \Omega_{DM}(z) \la{ode}
\end{eqnarray}
where $t(z)$ is as given in Eq~\ref{tr_reln}.
Notice that in (\ref{odm}) and (\ref{ode}) $t_0$ denotes (present) age of the 
universe. We plot $\Omega_{DM}(z)$ and $\Omega_{DE}(z)$ in fig (\ref{omega_dm_de}).
\begin{figure}[tbh!]
\includegraphics[width=7cm,height=7cm,angle=0]{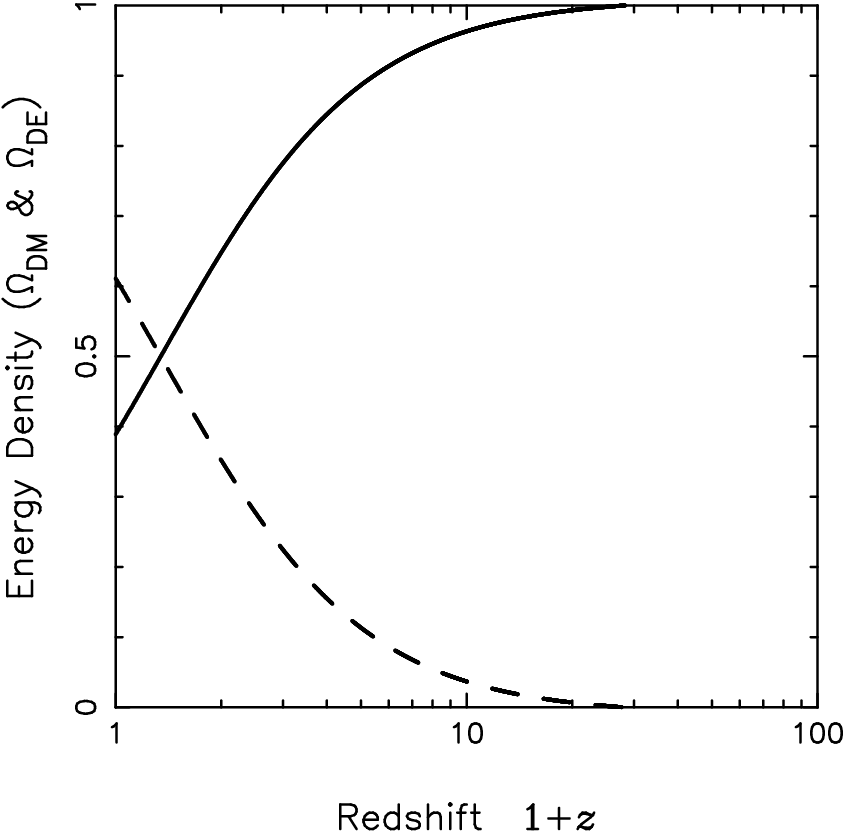} \caption{
This plot shows how $\Omega_{DM}$ and $\Omega_{DE}$ evolve with redshift
in our best fit model. The universe is completely dominated by matter
beyond a redshift $z \sim 10$. } \la{omega_dm_de}
\end{figure}

One way to test the validity of our model is to calculate its prediction of
$\Omega_{DM}$ and compare with the known value (in the LCDM model). It should
be noted that in this model these energy densities are only 'effective' energy
densities, and these components do not truly behave like the usual dark matter
and dark energy. However, since we can think of these densities as the
appropriate terms in the Taylor expansion of the Hubble parameter in our
model, it is likely that the local behaviour that fits the supernova data
would be similar, and hence these densities might compare well.  Another
constraint that our model must satisfy is the age of the universe,
which is known from independent measurements.  
One can define the so--called Hubble free luminosity $d_L$ distance using
$D_L = c H_0^{-1} d_L$. 
Using the definition of luminosity distance
\be
d_L(z)=(1+z)\int_{0}^{z}dz^{\prime}\frac{H_0}{H(z^{\prime})}\,\,,
\ee
we derive the expression for the same in our model as:
\begin{equation}
d_{L} \left (z ; \frac{p}{r}, s \right ) = \frac{(1+z)}{(1-\frac{p}{r})}
\int_{0}^{z} \left \{1- \frac{p}{r} \left [\frac{(1+z^\prime)
{\frac{1}{s}}}{1-\frac{p}{r} +\frac{p}{r} (1+z^\prime)^{\frac{1}{s}}}\right ]
\right \} dz' \,\,.
\end{equation}

In order to test the suitability of the present model 
of a tachyon condensate as the origin of dark matter 
and dark energy we have used the observational Gold dataset comprising
$157$ Type~Ia supernovae {\cite{golddata}}. The Hubble
constant free distance modulus is given by
\begin{equation}
\mu(z)=   5\,\log_{10}\,\left( d_L(z)/{\rm Mpc}\right) + 25\,\,.  
\label{eq:distance_mod}
\end{equation}
For model fitting we need to consider the likelihood function
given by
\ba
{\cal L} \propto \exp \left ( -\frac{\chi^2\left (H_0,p/r,s \right)}{2} \right ), \\
\chi^2 = \sum_{i=1}^{157} \left( \frac{\mu_i-\mu_{\rm fit}(H_0,p/r,s,z_i)}{\sigma_i} \right)^2 \,,
\ea
where $\mu_i$ and $z_i$ are the observed distance modulus and the redshift, and $\mu_{\rm fit}$ is
the expression Eq~\ref{eq:distance_mod} with the luminosity distance given
in Eq~21. The posterior probability for the model parameters $(H_0, p/r, s)$ is
given by
\be
{\rm P}(H_0, p/r,s) \propto \exp \left (
-\frac{\chi^2}{2} \right ) {\rm Pr}(H_0) \,,
\ee
where ${\rm Pr}(H_0)$ refers to the prior on the Hubble parameter. For our analysis we
have used the Gaussian prior, $ H_0= 66\pm 6 $ ({\cite{tegmark03}}).  
We need to define the bounds for the 
three-dimensional volume in $(H_0,  p/r, s)$. The lower and upper bounds
on $H_0$ are taken at  $ 50$ and $80 $ . It is obvious from the following
expression that $p/r<1$.
\begin{equation}
H_0 = \frac{1}{C(1-p/r)}\,\,.
\end{equation}
 But for q there is no upper bound. So we choose a
sufficient range for these parameters ($0.1<p/r<0.91$ and $0.66<s$)
in such a way that $p/r \times s$ is of the order unity.  

Starting from the best-fit $p/r,s$ (the value at the maximum of the
probability distribution), we may move down till $68
\%$ probability is enclosed under the surface and obtain the 
$1\sigma$ bound on the parameters. The $2 \sigma$ and  $3 \sigma$ bounds can be 
similarly obtained.  In Fig \ref{upperrange}.
we plot the $68$ \% (solid), $90$ \%(broken) and $ 95$ \% (dotted) contours.  The maximum of the
likelihood surface is at $s=0.66$ and
$p/r=0.385$. The $\chi^2_{\rm min}/{\rm d. o. f.}$ for this fit is $ 1.16$.
We simultaneously obtain $H_0=64.00$.  One can see that the contours are not
closed in the allowed region in the parameter space. 
For this fit, The value of $C \alpha = 0.76$ which satisfies the constraint 
stated earlier. In Fig~\ref{hubbleplot} we compare the present fit with
the canonical best fit $\Lambda CDM$ model.
\begin{figure}[tbh!]
\includegraphics[width=7cm,height=7cm,angle=0]{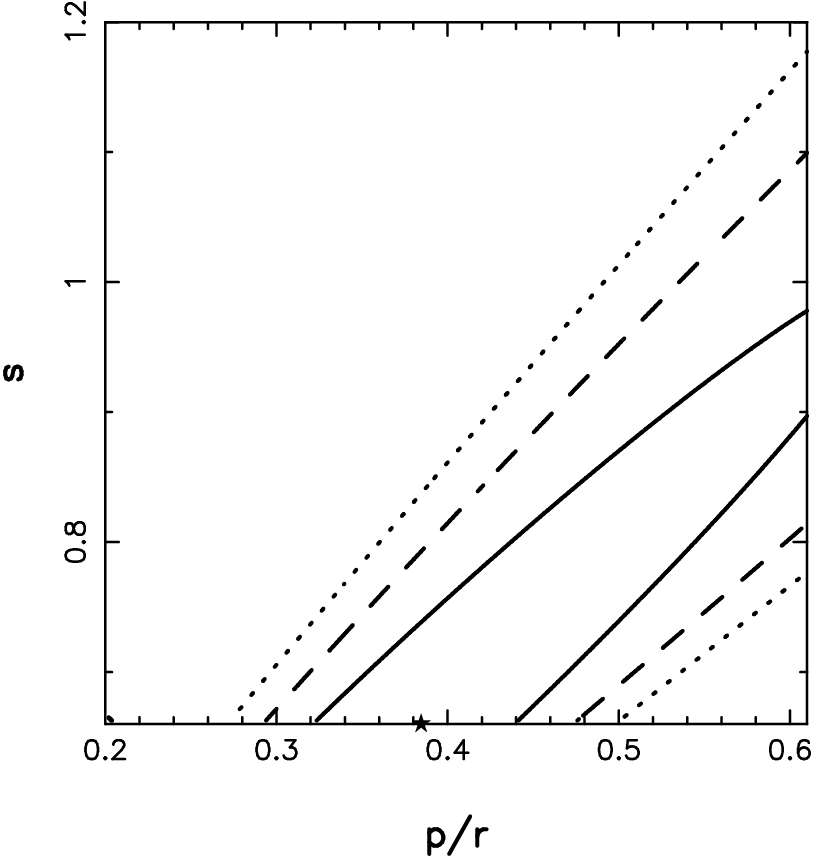}
\caption{The innermost, middle and the outermost
  contours correspond to $68\%, 90\%$ and $95\%$ confidence level
  respectively. Peak of likelihood surface corresponds to p/r=0.385, s=0.66.
  The asterisk marks the minimum of $\chi^2$.  As discussed in the text, to
  keep the potential physical at all times we must impose the constraint $s
  \ge 2/3$. We find that the maximum of the probability density lies at the
  edge of this interval.
}\label{upperrange}
\end{figure}

\begin{figure}[tbh!]
  \includegraphics[width=7cm,height=7cm,angle=270]{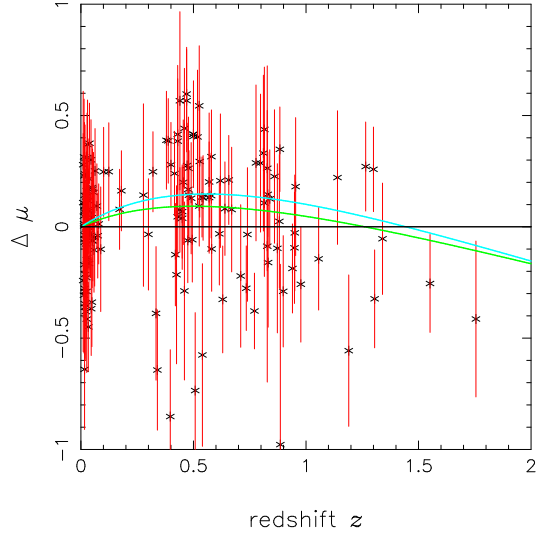} 
\caption{
  The green curve is the Hubble plot for the best fit model with $p/r=0.385$,
  $s=0.66$ and $H_0=64.0$. For comparison we also show the
  best fit $\Lambda CDM$ model (blue curve). 
}
\label{hubbleplot}
\end{figure}

Accepting the values of the parameters obtained from fit we can now estimate
various properties of the universe and compare with known values. Since best
fit $p/r=0.385, s=0.66, H_0=64.0$ and we argued $p=1$, it follows, $r=2.6$ and $\alpha=0.46H_0$ . Thus the age of the universe
according to this model is $t_0=\log ({r})/{2\alpha}=15.2$ Gyr.  Also
$(\Omega_{DM})_{0}=0.39, (\Omega_{DE})_{0}=0.61$, $q_0=p/rs -1=-0.42$.  These
values are close to the currently accepted values. The deceleration parameter
and redshift in this model are related by 
\be
q(z)=\frac{0.58}{0.615(1+z)^{-1.5} + 0.385} -1\,\,,
\ee 
and is found to represent
an universe which is accelerating at present epoch but was decelerating in
remote past (see fig (\ref{dec_param})). This fact is also in agreement with
current picture. The transition takes place at $z=1.15$. Note also that apart
from the best--fit values there are a whole range of values of the parameters
which are allowed at different confidence levels.
                
\begin{figure}[tbh!]
\includegraphics[width=7cm,height=7cm,angle=0]{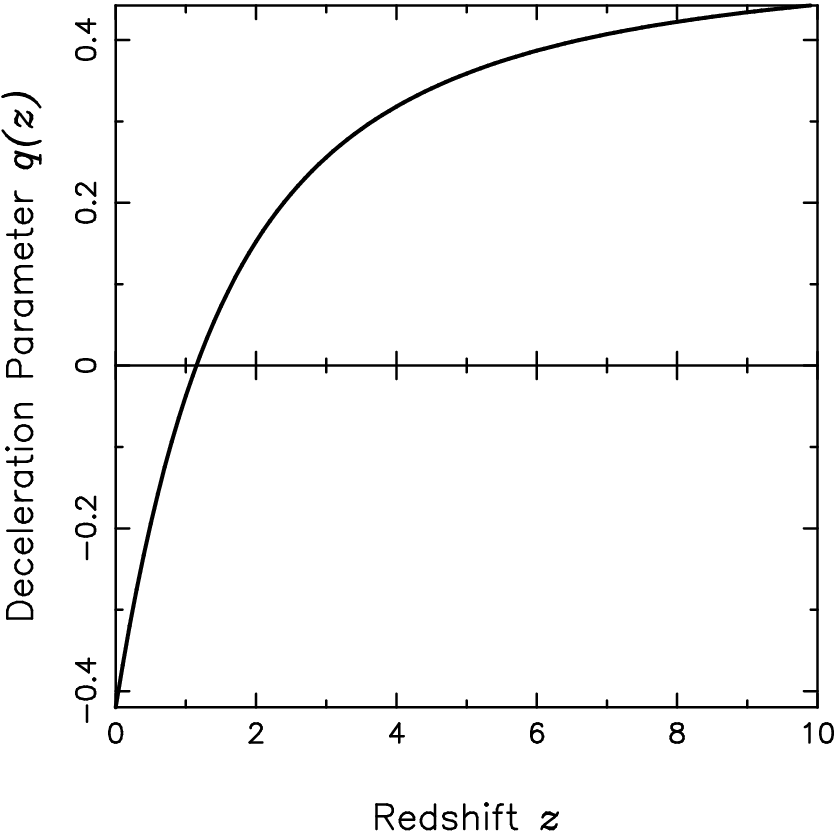} \caption{
The deceleration parameter calculated for the best fit model as in previous
figure. We see that the universe starts decelerating around $z \sim 1$.
}
\label{dec_param}
\end{figure}

It is obvious that there are many inadequacies in this model. For instance
in the remote past there should be a radiation--dominated phase and the
amount of radiation should decay to very small values in the present epoch.
At best, this model can accomodate the late--time accelerated phase 
which, by and large, has been our goal. We have an analytic expression
for the scale factor and the tachyon field. In addition, we have
made a simplistic assumption about dark matter and dark energy being
born out of the same scalar field. In some sense, our model tests how good 
this assumption is. The fact that the model fits the data quite well
is appealing but to improve on it we must add further necessary
features which are required for a realistic cosmological model. 
Finally, it might
be worthwhile looking at cosmological perturbations in this model and
try to use WMAP data to set more stringent bounds on the parameters {\cite{
tegmark03, WMAP}.

\begin{acknowledgments}
  Work of AD and SG is supported by a Junior Research Fellowship by the
  Council of Scientific and Industrial Research, India.
\end{acknowledgments}

\end{document}